# Direct observation of nanometer-scale amorphous layers and oxide crystallites at grain boundaries in polycrystalline $Sr_{1-x}K_xFe_2As_2$ superconductors

Lei Wang, Yanwei Ma[a]

Key Laboratory of Applied Superconductivity, Institute of Electrical Engineering, Chinese

Academy of Sciences, P. O. Box 2703, Beijing 100190, China

Qingxiao Wang, Kun Li and Xixiang Zhang[a]

Imaging and Characterization Core Lab, King Abdullah University of Science and Technology,

Thuwal 23955-6900, Saudi Arabia

Yanpeng Qi, Zhaoshun Gao, Xianping Zhang, Dongliang Wang, Chao Yao, Chunlei Wang

Key Laboratory of Applied Superconductivity, Institute of Electrical Engineering, Chinese

Academy of Sciences, P. O. Box 2703, Beijing 100190, China

## Abstract

We report here an atomic resolution study of the structure and composition of the grain boundaries in polycrystalline $Sr_{0.6}K_{0.4}Fe_2As_2$ superconductor. A large fraction of grain boundaries contain amorphous layers larger than the coherence length, while some others contain nanometer-scale crystallites sandwiched in between amorphous layers. We also find that there is significant oxygen enrichment at the grain boundaries. Such results explain the relatively low transport critical current density ($J_c$) of polycrystalline samples with respect to that of bicrystal films.

[a] Author to whom correspondence should be addressed; E-mail: ywma@mail.iee.ac.cn and xixiang.zhang@kaust.edu.sa

The discovery of superconductivity in Fe-As based compounds, with a critical transition temperature as high as 55 K, has sparked enormous research effort to understand its structure-property relationships [1-6]. One aspect of Fe-As based superconductors that has particular technological importance is that the grain boundaries in bicrystals do not appear to lower the overall critical current ($J_c$) as heavily as $YBa_2Cu_3O_{1-\delta}$ [7, 8]. A remarkable transport inter-grain $J_c$ of ~$10^5$ A/cm$^2$ at 4 K and self field has been reported in [001]-tilt cobalt-doped $BaFe_2As_2$ bicrystals with a high misorientation angle of 45$^o$ [8], offering the possibility of a number of industrial applications. This is in contrast to polycrystalline samples, which show a relative low inter-grain (transport) $J_c$ of ~$10^3$ A/cm$^2$ at 4.2 K and self field [3-5, 9]. The first critical step in understanding the mechanism by which polycrystalline samples can not support a high transport $J_c$ is to determine the structure and composition of the grain boundaries.

In this paper, we report a series of atomic resolution studies of the grain boundaries in a polycrystalline $Sr_{0.6}K_{0.4}Fe_2As_2$ sample. Three distinctive types of grain boundaries are found: clean grain boundaries, boundaries containing amorphous layer with a width of ~10 nm (i.e., larger than the coherence length), and boundaries containing amorphous-crystallite-amorphous trilayers ~30 nm in width. Furthermore, there is significant oxygen enrichment in the amorphous layer and crystallites. Such boundary features indicate that a large fraction of grain boundaries act as critical-current barriers, and that control of oxygen content at grain boundaries will be essential for attaining optimal polycrystalline critical currents.

The polycrystalline $Sr_{0.6}K_{0.4}Fe_2As_2$ samples used in this study were synthesized by one-step solid state reaction, which produces a sharp critical temperature ($T_c$) transition at ~35 K and an quasi single phase (for a more detailed description of the sample preparation see Ref. 10). One way to directly observe structure, composition, and bonding effects at grain boundaries is through the combination of high-resolution transmission electron microscopy (TEM) and electron energy loss spectroscopy (EELS) at atomic resolution in a scanning transmission electron microscope (STEM). Our experiments here use an FEI Titan 80-300 kV(ST) S/TEM operated at 300 kV which has a point resolution of ~0.205 nm in TEM mode and a 0.14 nm STEM resolution. Under this experimental set-up, the high-resolution grain boundaries structure can be observed directly at TEM mode, while STEM imaging can be used to position the electron probe for EELS. .

Fig. 1a shows a grain boundary network in polycrystalline $Sr_{0.6}K_{0.4}Fe_2As_2$ at low magnification. The $Sr_{0.6}K_{0.4}Fe_2As_2$ grains, with typical grain sizes around 5μm, appear to be well connected. Fig. 1b shows a high-resolution TEM image of a typical, clean high-angle grain boundary (Type-A), in which the sample was tilted so that the grain boundary was almost parallel to the incident electron beam. Only the (002) $Sr_{0.6}K_{0.4}Fe_2As_2$ planes ($d_{002}$=0.658nm) are resolved in the both grains. The selected area diffraction pattern of the across section of the grain-boundary region is shown in the inset of Fig.1b. The c-axis of one $Sr_{0.6}K_{0.4}Fe_2As_2$ grain is rotated by about $60^o$ with respect to that of the other.

It is interesting to note that some apparently well connected grains are not actually well connected. Fig. 1c shows the detailed structure of a grain boundary between two apparently well connected grains. It contains an amorphous layer of about 10 nm in thickness (Type-B), which is lager than the coherence length ~1-2 nm [11-12], consequently resulting in a current blocking effect. Similar amorphous layers around individual grains have been reported in $NdFeAsO_{1-\delta}$ and $Sr_{0.6}K_{0.4}Fe_2As_2$ [13-14]. Another type of grain boundary labeled Type-C contains nanometer-scale crystallites sandwiched between amorphous layers, as shown in Fig. 1d. The nanometer-scale crystallites are confirmed to be an impurity phase from the selected area diffraction pattern (inset of Fig. 1d).

Fig. 2b and Fig. 3b show high-angle annular dark field (HAADF) line scans across Type-B (Fig. 2a) and Type-C (Fig. 3a) grain boundaries, respectively. The dimension of the grain boundaries determined by HAADF line scans are consistent with that obtain from high-resolution TEM images. Energy dispersive X-ray spectroscopy (EDS) - and EELS- line scans have been performed simultaneously across the grain boundaries to analyze the composition of amorphous layers and the unknown crystallites. Peaks of Sr, K, Fe, As and O have been detected by EDS, as shown in Fig.

2c and Fig. 3c. The ratio Sr:K:Fe:As in grain is 0.55:0.44:2.0:2.0, nicely consistent with nominal composition. The composition variations across the Type-B grain boundary are shown in Fig. 2d, which shows that the amorphous layer do not contain as much level of arsenic as in grains. To determine the oxygen and iron contents precisely, EELS-line scan was also performed. A very small probe size (0.3 nm) and a pixel step of 2 nm have been used to record the oxygen K, iron L edge (Fig. 2e). The corresponding composition variations strongly suggest that oxygen at the region of Type-B grain boundary is very high (Fig. 2f).

A different situation is found at Type-C grain boundaries where nanometer-scale crystallites are sandwiched between amorphous layers. Similar to the Type-B boundary described above, a low level of arsenic as well as a high level of oxygen is observed at the boundaries. However, as can be seen from the EDS- and EELS- line scan spectra in Fig. 3d, the composition of amorphous layers in Type-C grain boundaries are not consistent with that of Type-B grain boundaries. It shows a significant enrichment of strontium in the amorphous layers in Type-C grain boundaries. The crystallites sandwiched between amorphous layers show enormous changes in the EELS spectra compared to the grain features. Here the iron L-edge intensity is reduced dramatically and the oxygen K-edge is increased by a large amount (Fig. 3f). It should be mentioned that the Sr peak seems drop a little bit, but Fe and As peak peaks drops more, so the Sr relative concentration is higher.

EDS analyses have been performed on the crystallite (Area 1), amorphous layer (Area 2) and grain (Area 3), as shown in Fig. 4. As the EDS signal counts, especially for O, is not enough for accurate quantification, the data here just want to show that the Sr and O are really high in the area 1 and 2. According to TEM image of the unknown crystallites, d-space like 6.5Å, 3.5 Å, 3 Å, 2.6 Å and 2 Å has been found. According to the X-ray diffraction (XRD) data base, the structure of the unknown crystallites may be close to one of the tetragonal $SrO_2$ structure. In area 3 which is far from the grain boundary, the O is only 1.4%. This means the main phase of the material is still $Sr_{0.6}K_{0.4}Fe_2As_2$ and O only exist close to the grain boundary.

The results described above clearly show that the grain boundaries in $Sr_{0.6}K_{0.4}Fe_2As_2$ exhibit a propensity for oxygen enrichment. At this stage it is not clear whether the two different types of dirty boundaries we observed are fundamentally different or simply different stages of oxidation. While, it is immediately clear that Type-B and Type-C grain boundaries act as inter-grain (transport) critical current barriers, and by itself may be enough to explain the relatively low transport critical current density ($J_c$) of polycrystalline samples with respect to that of bicrystal films. It also indicates that control of oxygen content at grain boundaries will be essential for attaining optimal polycrystalline critical currents.

Authors are indebted to H. H. Wen, R. Flukiger, H. Kumakura, S. X. Dou and T. Matsushita for useful comments. This work is partially supported by the National '973' Program (Grant No. 2011CBA00105), National Science Foundation of China (Grant No. 51025726) and Beijing Municipal Science and Technology Commission under Grant No. Z09010300820907.

**Captions**

Figure 1 (a) TEM image of an apparently well connected grain boundary network in polycrystalline $Sr_{0.6}K_{0.4}Fe_2As_2$. (b) A high-resolution TEM image of a typical, clean high-angle grain boundary. (c) The detailed structure of a grain boundary containing an amorphous layer about 10 nm in thickness. (d) Another type of grain boundary containing nanometer-scale impurity crystallites sandwiched between amorphous layers.

Figure 2 (a) A STEM image of the Type-B grain boundary. (b) HAADF-line scan across the grain boundary. (c, d) EDS spectra and line scan performed on the amorphous layer. (e, f) EELS spectra and line scan performed on the amorphous layer.

Figure 3 (a) A STEM image of the Type-C grain boundary. (b) HAADF-line scan across the grain boundary. (c) EDS spectra and line scan performed on the amorphous-crystallite-amorphous trilayer. (e) EELS spectra and line scan performed on the amorphous- crystallite-amorphous trilayer.

Figure 4 EDS analysis on the crystallite (Area 1), amorphous layer (Area 2) and grain (Area 3).

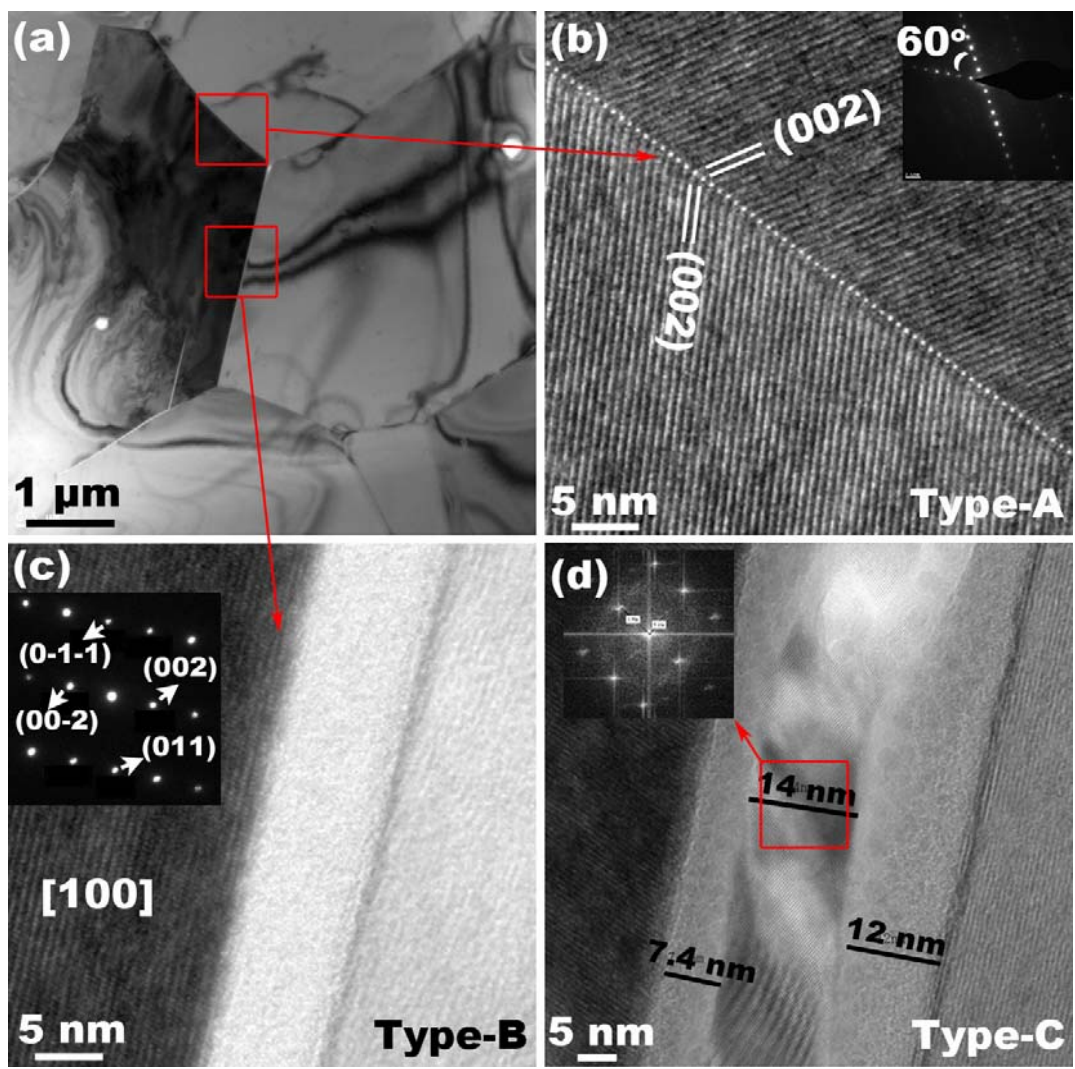

Figure 1 Wang et al.

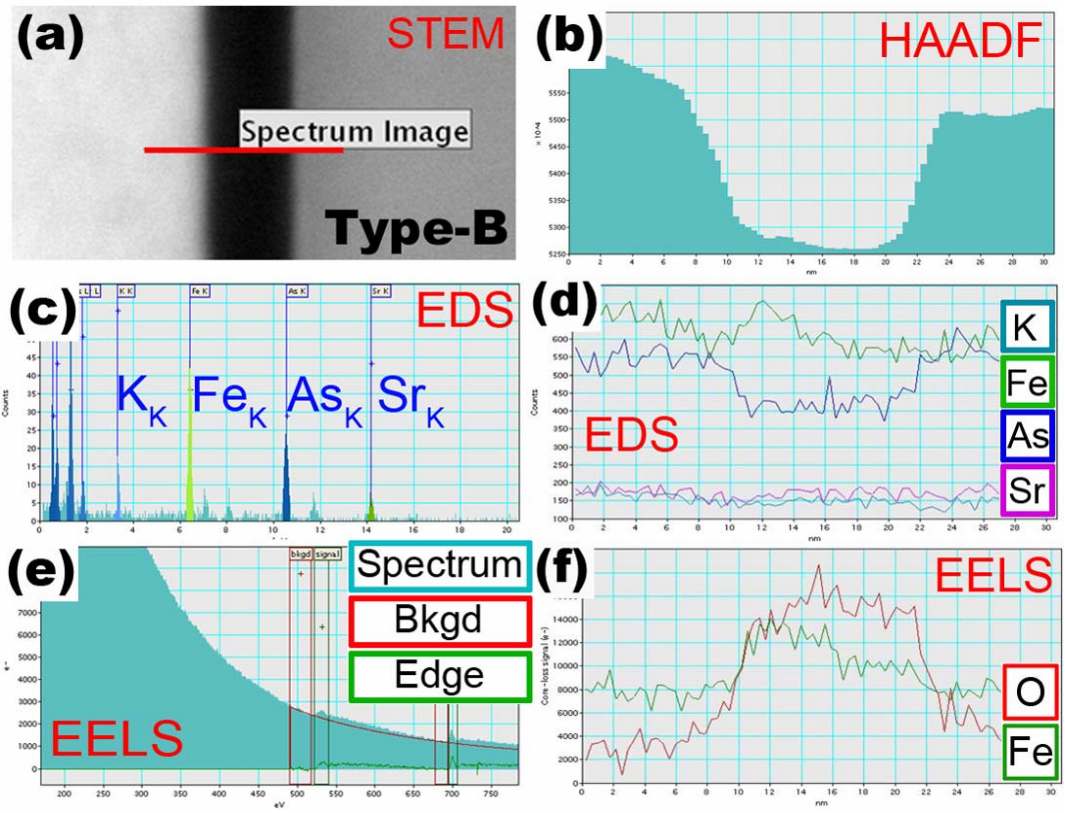

Figure 2 Wang et al.

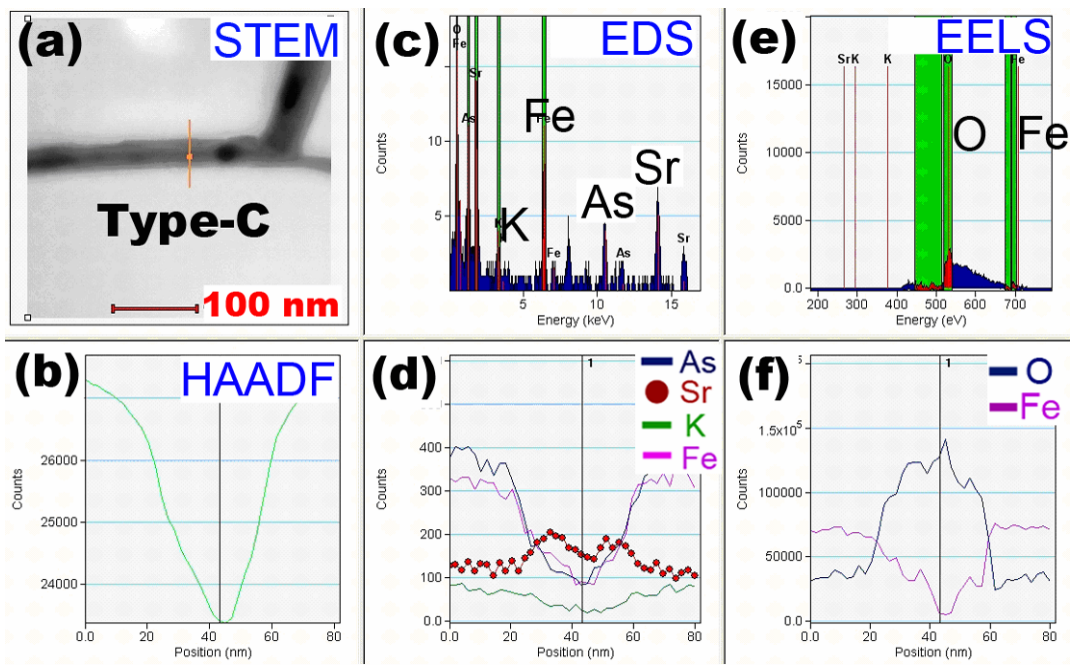

Figure 3 Wang et al.

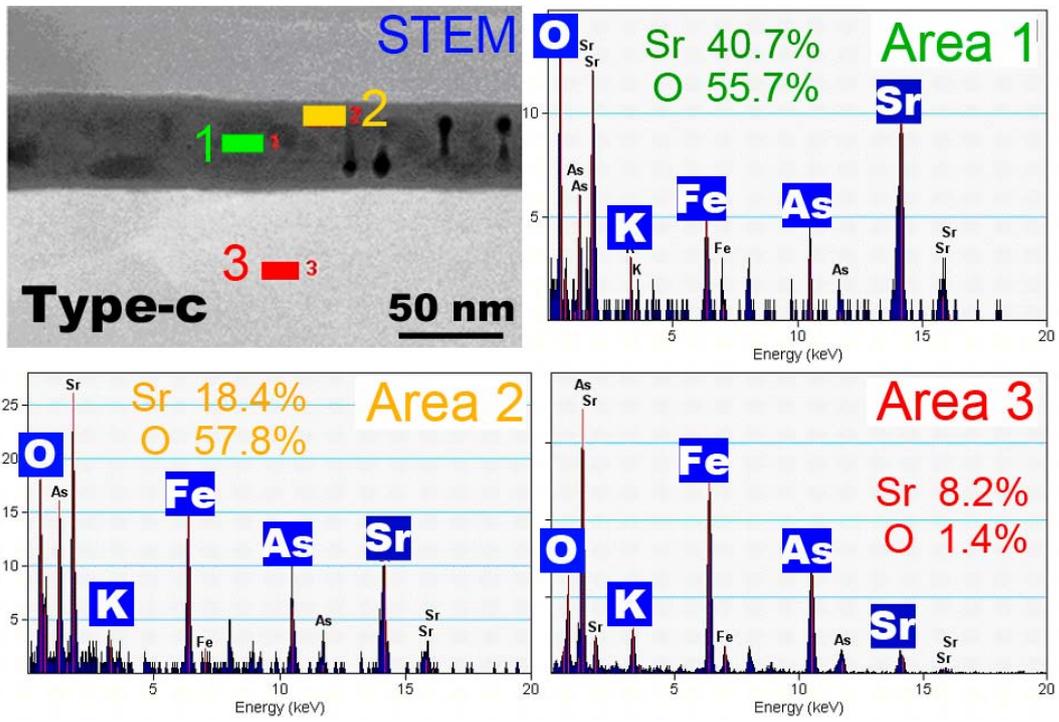

Figure 4 Wang et al.